\documentclass[12pt,preprint]{aastex}

\slugcomment{Submitted to the Astrophysical Journal Letters.}

\shorttitle{The gradual slow-down of powerful extragalactic jets}
\shortauthors{Georganopoulos \& Kazanas}

\begin{document}

\title{ Witnessing the gradual slow-down of powerful extragalactic jets: The X-ray -- optical -- radio connection.
}

\author{Markos Georganopoulos\altaffilmark{1}
 \& Demosthenes Kazanas\altaffilmark{2}}
\affil{Laboratory for High Energy Astrophysics, NASA Goddard Space Flight Center, 
Code 661, Greenbelt, MD 20771, USA.}
\altaffiltext{1}{Also NAS/NRC Research Associate; email:
markos@milkyway.gsfc.nasa.gov}
\altaffiltext{2}{email: Demos.Kazanas-1@nasa.gov}


\begin{abstract}

A puzzling feature of the {\it Chandra}--detected quasar jets  is 
that their X-ray emission decreases faster along the jet than
their radio emission, resulting to an outward increasing  radio to X-ray 
ratio. In some sources this behavior is so extreme that the radio 
emission peak is located clearly downstream of that of the X-rays.
This is a rather unanticipated behavior given that the inverse
 Compton nature of the X-rays  and the synchrotron radio emission
are  attributed to roughly the same electrons of the jet's non-thermal
electron distribution. In this note we show that this morphological 
behavior can result from the gradual 
deceleration of a  relativistic flow and that the offsets in peak emission 
at different wavelengths carry the imprint of this deceleration. This 
notion is consistent with another recent finding, namely  that the 
jets feeding the terminal hot spots of powerful radio galaxies and quasars
are still relativistic with Lorentz 
factors $\Gamma \sim 2-3$. The picture of the kinematics of powerful jets
 emerging from these considerations is that they remain
relativistic as they  gradually decelerate from Kpc scales to the hot spots,
where, in a final collision with the intergalactic medium, they slow-down
rapidly to the subrelativistic velocities of the hot spot advance speed.

\end{abstract}


\section{Introduction}

The superior angular resolution and sensitivity of {\it Chandra} has
led to the discovery of X-ray emission from a number of 
quasar jets. Schwartz et al. (2000) were the 
first to note that the X-ray emission  from the knots of the jet 
of the superluminal quasar  PKS 0637-752 (Chartas et al. 2001), 
at a projected distance $\sim 100$ Kpc from the quasar core,
is part of a spectral component separate from its synchrotron 
radio-optical emission and it is too bright to be explained through 
Synchrotron - Self Compton (SSC) emission from electrons in energy 
equipartition with the jet magnetic field (note however that  in the 
innermost knot  of some sources -- e.g. 3C 273 (Marshall et al. 2001), 
PKS 1136-165  (Sambruna et al. 2002) -- a synchrotron X-ray contribution is 
possible). These properties are apparently common to 
other quasars jets, as indicated by the mounting observational 
evidence (Sambruna et al 2001; Marshall et al. 2001; Jester 
et al. 2002; Sambruna et al. 2002; Siemiginowska et al. 2002; 
Jorstad, Marscher,  \& McHardy 2003; Siemiginowska et al. 2003; Yuan et al. 
2003; Cheung 2004,  Sambruna et al. 2004).

To account for the level of the observed  X-ray emission,
 Tavecchio et al. (2000) and Celotti et al. (2001) 
proposed that this is due to  External Compton (EC) scattering 
of the Cosmic Microwave Background (CMB) photons off relativistic 
electrons in the jet, provided that the jet flow is sufficiently 
relativistic ($\Gamma \sim 10$) to  boost  the 
CMB energy density in the flow frame (by $\Gamma^2$)  at the level 
needed  to reproduce  the observed X-ray flux.

In all these sources the radio-to-X-ray ratio  
increases downstream along the jet, an unexpected behavior, 
given that the cooling length of the EC X-ray emitting electrons 
($\gamma \sim$ a few hundreds) is longer than that of the radio 
emitting ones ($\gamma \sim$ a few thousands) and comparable or  longer than
the size of the jet, which would lead to a constant X-ray brightness 
as far out as the hot spots, contrary to observations. More surprisingly, 
in some jets 
(e.g. 3C 273 in Sambruna et al. 2001 and  Marshall et al. 2001; PKS 1136-135 and 1354+195  in Sambruna et al. 2002; PKS 1127-145 in Siemiginowska et al. 2002; 0827+243 in Jorstad, Marscher,  \& McHardy 2003)
the X-ray and radio maps are anti-correlated, with the 
X-ray emission peaking closer to the core, gradually decreasing 
outward, and the radio emission increasing outward to peak practically 
at the very end of the radio jet. 

To explain the reduction of the X-ray flux along the jet Tavecchio, 
Ghisellini \& Celotti (2003) suggested that the X-ray emission 
originates in a collection of micro-knots undergoing  adiabatic 
expansion sufficient to produce the desirable electron cooling. 
However, this would suppress also the radio emission, leading 
to practically indistinguishable radio and X-ray morphologies, 
contrary to their observed spatial anti-correlation. An elegant suggestion 
by Dermer \& Atoyan (2002) that the X-rays are  synchrotron emission 
from electrons cooling in the Klein--Nishina regime naturally produces 
shorter sizes  in X-rays than in radio. However, it  also  produces 
larger optical than X-ray sizes, contrary to observations, and it 
seems therefore not to be applicable in this particular context.

Our view is that, although X-ray producing electrons are present 
throughout the jet, the X-ray brightness decreases because the CMB 
photon energy density in the flow  frame decreases along
the jet as a result of a decelerating relativistic jet flow. The
decrease in the flow Lorentz factor leads to a decrease in the 
comoving CMB photon energy density and hence to a decrease in the X-ray 
brightness along the jet.
At the same time, the flow deceleration leads to a compression 
of the magnetic field; this results to an enhanced radio
emission with distance, which gets thus displaced 
downstream of the EC X-ray emission.

Based on the radio, optical, and X-ray jet maps, we argue 
in this note, that powerful extragalactic jets
are relativistic and gradually decelerating.
In \S 2 we formulate the synchrotron and EC emission 
process from a decelerating jet flow and present our results.
 In \S 3 
we discuss our findings and  touch upon some  open issues.


\section{The Model}

We parameterize the jet flow assuming that its radius $r$ 
scales as $r(z)=r_0 f(z)$, where $f(z)$ is a function with $f(z_0)=1$ 
and $r_0$ is the jet radius
at a fiducial point $z_0$ along the jet. Similarly, we assume that the flow
decelerates as  $\Gamma(z)=\Gamma_0 g(z)$,
where $g(z)$ is a function with $g(z_0)=1$ and  $\Gamma_0$ is
the bulk Lorentz factor at $z_0$. 
The evolution of the electron energy distribution  (EED) along the jet is determined by the combination
of adiabatic changes (losses for expansion and gains for compression), radiative losses, and particle
acceleration. If the radiative losses of the electrons responsible for the X-ray and radio emission are negligible 
(as may be the case for the ($\gamma \sim$ a few hundreds) EC X-ray emitting electrons, or if we assume that some localized (e.g. shocks; Kirk et al. 2000) or spatially distributed  (e.g. turbulence; Manolakou, Anastasiadis, \& Vlahos 1999) particle acceleration mechanism  offsets the radiative losses, then  
the evolution of the  EED can be treated as adiabatic. 
In this case, assuming that $\Gamma \gg 1$,
the elementary volume $dV$ scales as $dV=dV_0 f^2 g$, allowing for 
expansion or compression, while the adiabatic electron 
energy change rate with $z$ is obtained from  
${\gamma'}=-\gamma\left(2{f'/ f}+{g'/ g}\right)/3$,
 where the prime denotes differentiation with respect to $z$. 
The solution of the above equation is $\gamma(z)=\gamma_0f^{-2/3}g^{-1/3}$.
Assuming  particle conservation, $n(\gamma,z)dVd\gamma=n(\gamma_0,z_0)dV_0d\gamma_0$, 
the comoving EED   $n(\gamma,z)$ can be written as 
\begin{equation}
n(\gamma,z)=k \gamma^{-s}f^{-2(s+2)/3}g^{-(s+2)/3},
\label{n}
\end{equation} 
where $n(\gamma,z_0)=k\gamma^{-s}$ is the EED at $z_0$.
Following Georganopoulos, Kirk \& Mastichiadis (2001),
 the EED  in the local rest frame  is
$n(\gamma,z)\Delta^{2+s}$, where $\Delta(z)$
is the familiar 
Doppler factor $\Delta(z)=1/(\Gamma(z)(1-\beta(z)\cos\theta))$
and $\theta$ is the observing angle.
 Using this, and taking into account the cosmological corrections,
 the EC flux  per $dz$  in the $\delta$-function approximation 
 (Coppi \& Blandford 1990; an electron of Lorentz factor $\gamma$ upscatters seed photons of energy $\epsilon_0$ only to an    energy   $4\epsilon_0\gamma^2/3$)
 is 
\begin{equation}
{dF_c \over d\epsilon  dz}= { k \sigma_\tau c U (1+Z)^2 \over 2 d^2 \epsilon_0}             f^{-2(s-1)/3}g^{-(s+2)/3}\Delta^{2+s} \left({3\epsilon \over 4\epsilon_0}\right)^{-(s-1)/2},
\label{ec}
\end{equation}
where $Z$ and $d$ are  the redshift and proper distance of the source, $\epsilon_0$ and $\epsilon$ are 
respectively the energy of the CMB seed photons and the observed
photons in units of $m_ec^2$ at $Z=0$, and $U$ is the CMB energy density 
at $Z=0$. For $\theta=0$, $\Delta(z)=2\Gamma(z)$ and
the $z$-dependence of the flux   scales  as $\propto     f^{-2(s-1)/3}g^{(4+2s)/3}$.

Assuming flux conservation, the magnetic field $B(z)$ can be written as $B(z)=B_0/fg$. The synchrotron flux   per $dz$ in the $\delta$-function approximation (an electron of Lorentz factor $\gamma$ produces synchrotron photons only
at  the critical synchrotron 
energy $B\gamma^2/B_{cr}$) is then
\begin{equation}
{dF_s \over d\epsilon  dz}={k \sigma_\tau c B_{cr} B_0^{(s+1)/2}\over 12 \pi d^2 (1+Z)^{(s+1)/2}} f^{-(7s-1)/6}g^{-(5s+7)/6}\Delta^{(s+3)/2} (\epsilon B_{cr})^{-(s-1)/2},
\label{s}
\end{equation}
where $B_{cr}=m_e^2 c^3/e\hbar=4.414 \; 10^{13}$ G is the critical
 magnetic field.
 Note that while the synchrotron flux decreases with increasing
redshift  (the quantity
 $d^2 (1+Z)^{(s+1)/2}$ increases), the EC flux ($\propto (1+Z)^2/d^2$) 
after an initial decrease
up to $Z\approx 1.9$, remains practically constant for higher $Z$, as
was first noted by Schwartz (2002).  
 
For $\theta=0$ the 
$z$-dependence of the flux   scales as  $\propto   f^{-(7s-1)/6}g^{-(s-1)/3}$.
The condition that at $\theta=0$ the synchroton flux
increases along the jet is that deceleration dominates over jet opening,  
leading to $-{g' / g}<  (7s-1)f' / 2(s-1) f$.
Taking the ratio of equation (\ref{s}) to equation  (\ref{ec}), the condition
that  $L_s/L_{EC}$  (or equivalently the radio to X-ray spectral index $\alpha_{rx}$) increases with $z$ at $\theta=0$ is
$-g' / g<   f' / 2 f$,
and for a given jet opening profile $f$ a  milder  deceleration $g$
is needed to produce an   increase in $\alpha_{rx}$  than to separate
the X-ray and radio peaks. This is in agreement with observations that
show that the decline of the radio to X-ray flux is a more widely
observed trend that the X-ray -- radio peak displacement.

\begin{figure}
\epsscale{0.8}
\plotone{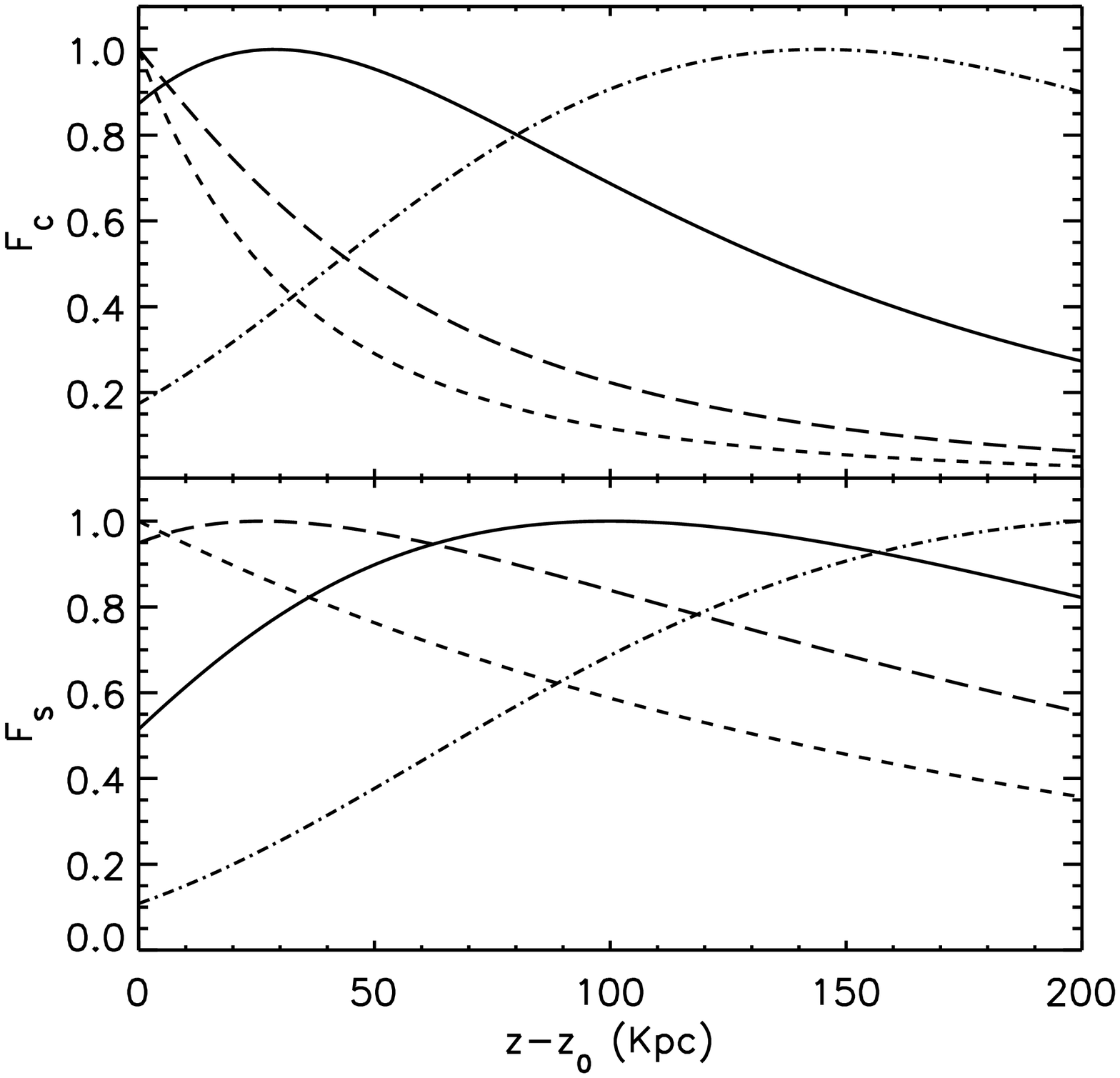}
\caption{The EC (upper panel) and synchrotron (lower panel) flux  normalized to its maximum value for a range of observing angles along a jet  that decelerates from
 $\Gamma_0=6$ to $\Gamma=2$ with $\epsilon=1$ from  $z_0=100$ Kpc to
 $z_{max}=300$ kpc, 
while its radius doubles. The observing angles are $\theta=0^\circ$ (short dash line),
$\theta=1/2\Gamma_0\approx 4.8^\circ$ (long dash line), $\theta=1/\Gamma_0\approx 9.6^\circ$ (solid line),              $\theta=2/\Gamma_0\approx 19.5^\circ$ (dot dash line).}
\label{fig1}
\end{figure}

 For the rest of this work
we consider  conical jets with $f=(1+az)/(1+az_0)$, where $a=0$ corresponds to a cylindrical flow,
and  deceleration profiles $g=(z/z_0)^{-\epsilon}, \; \epsilon \geq 0$.
The behavior of the EC and synchrotron flux  profiles depends also on the observing angle.
To demonstrate this, we plot in  
 the upper (lower) panel of Figure \ref{fig1} the EC (synchrotron) 
flux  along a decelerating jet, for a range of angles. 
It can be seen that, while for small angles the EC flux
peaks at the base of the jet, at larger angles the peak shifts 
 downstream due to the fact that the emission from
the faster base of the jet is beamed out of the line of sight
 (for a given $\theta$, $\Delta(z)$ peaks at $\Gamma(z)\approx 1/\sin \theta$).
The behavior of the synchrotron flux is also interesting.
 At  angles $\theta \not= 0$ 
the peak of the synchrotron emission shifts downstream and 
 for a given angle it is further downstream than the EC peak, in agreement with  observations.

It is interesting to see how much the above considerations change
when  radiative losses are important. 
The simplest case is that of a perfectly collimated
(i.e. $\alpha =0$)  cylindrical jet. This is also the case that
requires the mildest deceleration for producing a given 
$\alpha_{rx}$ increase or X-ray -- radio separation.
Assuming an   EED slope $s=2$, the EED remains a power law of 
the same slope  following equation (\ref{n}), with a  maximum energy electron Lorentz factor  
\begin{equation}
\gamma_{max}(z)={\gamma_{max(z_0)}z_\star^{\epsilon/3} \over 1+\gamma_{max(z_0)} [C_1 (z_\star^{(10\epsilon+3)/3}-1)+C_2(z_\star^{(3-2\epsilon)/3}-1)]},
\end{equation}
where $\gamma_{max(z_0)}$ is the upper cutoff of the EED at $z_0$,             $z_{\star}=z/z_0$, $C_1=3z_0AB_0^2/8\pi(10\epsilon+3)$,
$C_2=4z_0AU(1+Z)^4\Gamma_0^2/(3-2\epsilon)$, and $A=4\sigma_\tau/3m_ec^2\Gamma_0$. The maximum EC energy $\epsilon_{EC,max}$ as a function of $z$ is then 
 $\epsilon_{EC,max}=\epsilon_0 \Delta^2(z) \gamma_{max}^2(z)$, while
the maximum synchrotron energy  is 
$\epsilon_{s,max}=B_0z_{\star}^\epsilon \Delta(z)\gamma_{max}^2(z)/B_{cr}$.
For a given observing energy and observing angle, 
one can find the radiating  part of the jet
and perform the optically thin radiative transfer by integrating
expressions (\ref{ec}) and  (\ref{s}) along each line of sight for a two
dimensional array of lines of sight that covers the source as projected on
the observer's plane of the sky.

\begin{figure}
\epsscale{0.8}
\plotone{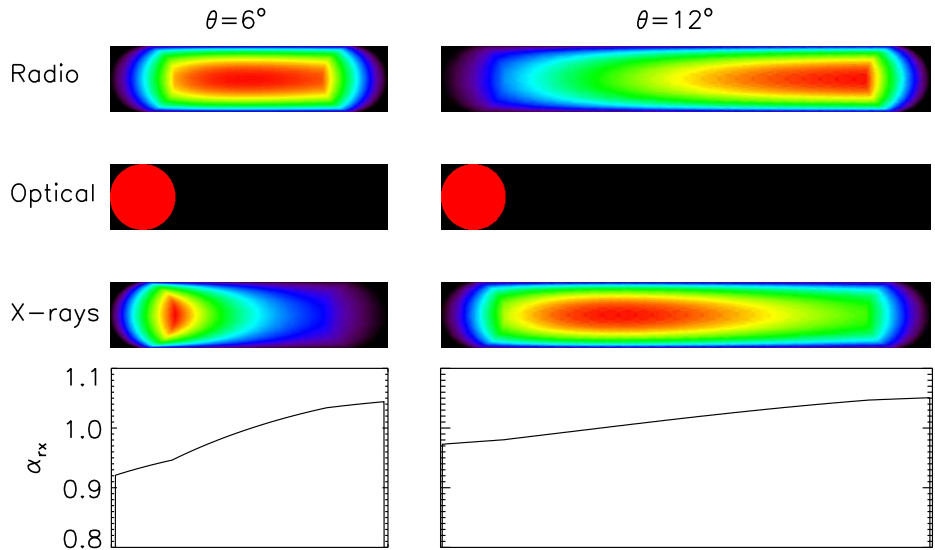}
\caption{The radio (top), optical(middle), X-ray (bottom) maps and the radio to X-ray  spectral index $\alpha_{rx}$ (below the X-ray map) for a decelerating flow
observed under $\theta=6^\circ$ (left) and $\theta=12^\circ$ (right). The cylindrical 
flow,  assumed to be at $Z=0.3$, decelerates from $\Gamma_0=6$ to $\Gamma=2$ 
with $\epsilon=1$ and $B_0=3.\,  10^{-6}$ as it propagates a distance of 200 Kpc.  The EED is a power-law of index $s=2$ and $\gamma_{max}=10^6$ at the base
of the flow. This calculation includes radiative losses. Each map is normalized with red indicating the maximum luminosity per unit area. The projected jet lengths are $\approx 21$ Kpc for $\theta=6^\circ $ and $\approx 42$ Kpc for            $\theta=12^\circ$.}
\label{fig2}
\end{figure}

 The results of such  calculations
are shown in Figure \ref{fig2}, where it can be seen that the optical emission is confined at the base of the flow,
due to the strong radiative losses of the high energy electrons, practically marking the site of strong particle acceleration. The morphology of the radio emission is angle
dependent: at $\theta=6^{\circ} $ it covers the entire extend of the  jet, while at  $\theta=12^{\circ} $ the emission of its inner
part is is dimmed because most of the radiation is beamed outside our line of sight. The X-ray emission at both
angles peaks close to the base of the flow, due to the decrease of the comoving CMB 
photon energy density away from the flow base.
Regarding the relative radio-X-ray morphology, while at $\theta=12^{\circ} $ the offset between the two images
is very large, at  $\theta=6^{\circ} $ the two images overlap, although  the radio image clearly  extends 
further downstream.   Also, as the radio-X-ray spectral index plots show, in both cases $\alpha_{rx}$ increases 
downstream in agreement with observations.  

\begin{figure}
\epsscale{0.8}
\plotone{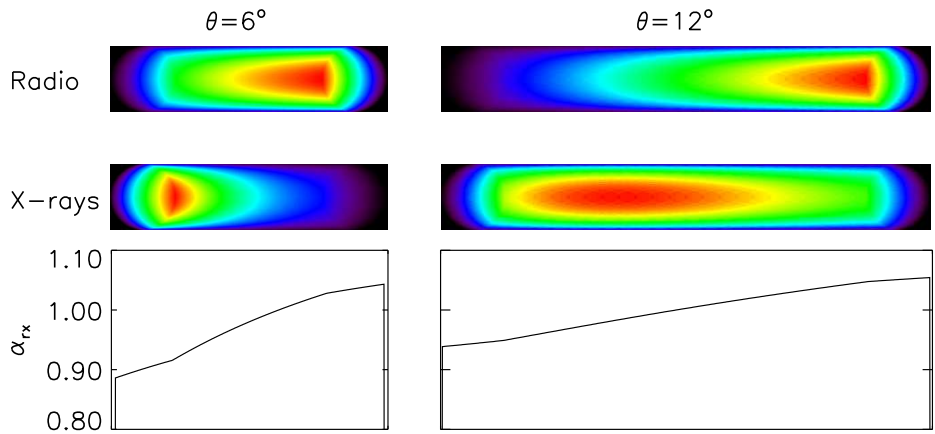}
\caption{The same as in figure (\ref{fig2}) but with radiative losses balanced by distributed reacceleration and without the optical map, which
in this case would be identical to the radio map.}
\label{fig3}
\end{figure}

We compare now this picture with the one we would get from the same system  if radiative losses were balanced by distributed reacceleration.
(Figure \ref{fig3}).  The X-ray images remain practically the same, because the radiative losses for the
 EC X-ray emitting electrons in this  $Z=0.3$, $\Gamma_0=6$ jet are negligible.
The radio image however changes significantly, particularly so for the $\theta=6^{\circ} $ orientation. Now
that the radio emitting electrons retain their energy the outer part of the jet is much brighter than 
the inner part, resulting to a clear separation between the radio and X-ray images.


\section{Discussion and Conclusions}

We have proposed that the increase of the radio-to-X-ray flux ratio
along the length of the jets of powerful quasars, as well as the 
occasional offset of the jet images in these wavelengths, are naturally 
accounted for in terms of relativistic flows that decelerate over the entire
length of the jet. Despite this deceleration, the jets remain relativistic  
($\Gamma \sim$ a few) to their terminal hot spots (Georganopoulos \& Kazanas 
2003; hereafter GK03), within which they eventually attain sub-relativistic 
speeds. Our proposal provides, for the first time, a means for deducing 
the jet kinematics through simple models of their multiwavelength images;
these can then be checked for consistency when coupled to the detailed 
hot spot emission, which as shown in earlier work (GK03), depends on 
the value of $\Gamma$ at this location.
We note here that a change of the Doppler factor can be also produced 
if the jet curves monotonically away from the line of sight without actually
decelerating. However, in this case the CMB comoving photon energy density would 
remain constant along the jet, and the X-ray and radio emission would
decrease along the jet in a similar manner, contrary to observations.

The scenario we propose here finds further support from modeling jets 
with  several sequential individual knots: Sambruna et al. (2001) 
noted the need for a gradual decrease of the Doppler factor and/or 
an increase of the magnetic field in order to reproduce the emission 
from the knots along the jet of 3C 273 with simple one zone models. These 
knot to knot variations can be naturally incorporated within the 
context of  a decelerating collimated flow, as we propose.

These same maps indicate also the need or not of distributed particle
reacceleration along these jets, when the EC loss length scale of their
radio emitting electrons (proportional 
to $(1+Z)^{-4} \Gamma^{-2}$) is shorter than the observed length of 
the radio jet. This appears to be the 
case with sources like PKS 1127-145 at $Z=1.187$ (Siemiginowska et al. 
2002) and possibly GB 1508+5714 at $Z=4.3$ (Yuan et al. 2003; 
Siemiginowska et al. 2003; Cheung 2004), which show their peak radio 
emission displaced from that of the X-rays: in the  absence of reacceleration, 
EC losses of the radio emitting electrons would produce radio jets shorter 
than the X-ray ones.  Much about the reacceleration
process at these jets can be inferred from their morphology at different  
frequencies. For example, the  knotty optical jet morphology shows that 
reacceleration  
is not strong enough to offset the EC-dominated losses of the optically 
emitting synchrotron  electrons ($\gamma \sim 10^{6-7}$). 

The radiative efficiency of these jets is less than a few percent (e.g. 
Tavecchio et al. 2000) and the energy lost in deceleration must be either 
used to heat up the matter in the jet, or must be transferred to material 
that is entrained by the jet. While the first option would result to
an expansion of the jet, contrary to what is seen, entrainment from an 
external medium would load the jet  with baryonic mass while decelerating 
it. 
 Entrainment would produce velocity gradients
across the jet, and in this sense a faster spine and a slower sheath
are to be expected. However, the observed X-ray and radio 
jet morphologies suggest that  the dominant effect must be a deceleration
 along the jet.
A fast spine that  does not decelerate substantially
  and carries most of the jet power would
produce  a constant X-ray EC and radio synchrotron flux along the jet, 
in disagreement
 with observations.  
A consequence of entrainment would be that even if the jet did not start as a 
baryonic one,  entrainment would gradually enrich it with baryons and 
eventually a fraction $\Delta\Gamma/\Gamma$ of its power, where  $\Delta
\Gamma$ is the decrease of the Lorentz factor, would be carried by the 
entrained baryonic matter.    
This in turn would increase the radio lobe equipartition energy content
 estimates derived under the assumption of a leptonic composition.
 Our findings point to a picture
where powerful relativistic jets decelerate, depositing  most of their power
 in their surroundings in the form of kinetic energy, their observed radiation
being  only the tip of the  iceberg.

\acknowledgments We want to thank the referee  for his/her suggestions.
G. M.  thanks Rita Sambruna and Jessica Gambill for illuminating discussions on recent jet observations.

\clearpage

\clearpage

\end{document}